\documentstyle[aps,twocolumn]{revtex}
\begin{document}

\title{Bose-Einstein Condensation in Geometrically Deformed Tubes}

\author{P. Exner$^{1,2}$ and V.A. Zagrebnov$^{3}$}
\address{
$^{1}$Nuclear Physics Institute, Czech Academy of Sciences,
25068 \v{R}e\v{z} near Prague, Czech Republic \\
$^{2}$Doppler Institute, Czech Technical University,
B\v{r}ehov\'{a} 7, 11519 Prague, Czech Republic \\
$^{3}$Universit\'e d'Aix-Marseille II and Centre de Physique
Th\'eorique, CNRS-Luminy-Case 907, 13288 Marseille, France}

\maketitle

\date{\today}

\begin{abstract}
We show that Bose-Einstein condensate can be created in
quasi-one-dimensional systems in a purely geometrical way, namely
by bending or other suitable deformation of a tube.
\end{abstract}

\narrowtext

\pacs{PACS: ????}

\vspace*{-5mm}

It is well known\cite{ZB} that there is no Bose-Einstein
condensate (BEC) of the continuous free Bose gas in one dimension
as well as in tube-shaped domains $D\times\mathbf{R}$ with a
compact $D\subset\mathbf{R}^2$. On the other hand, there is an
increasing experimental and theoretical interest \cite{interest}
to the BEC in quasi-one-dimensional systems of trapped boson
gases, in particular with the idea to understand properties
predicted by the Lieb-Liniger model \cite{LL} and the
Girardeau-Tonks gas \cite{GT}.

In this letter we want to draw attention to a way to create BEC in
a quasi-one dimensional perfect Bose gas (PBG) based on
\textit{geometrical} properties of the recipient \cite{ACB}.
Dealing with a PBG we have naturally to describe first the
one-particle spectrum. Consider first a \emph{straight}
round-tube-shaped recipient (cylinder) ${\cal C}(S_r) =
S_r\times\mathbf{R}$ of radius $r>0$ and the associated Dirichlet
Laplacian \cite{RS}
\begin{equation}\label{Lap}
  t := -\Delta_D^{{\cal C}(S_{r})},\quad
  {\rm Dom}(t) = W^{2,2}_0(S_{r}\times\mathbf{R})\,.
\end{equation}
The spectrum of $t$ equals $\bigcup_{j=1}^{\infty}\,[E_j,
\infty)$,
where  
\begin{equation}\label{2-spect}
0 < E_1 < E_2\leq E_3\leq\ldots
\end{equation}
are the eigenvalues of the two-dimensional operator $
-\,(\partial^{2}_x +\partial^{2}_y)$ with the Dirichlet boundary
conditions on the tube cross section $S_{r}$. The lowest one is,
of course, non-degenerate; for the circular cross section we have
$E_1= \frac{\hbar^2}{2M}\, j_{0,1}^2 r^{-2}$, where $j_{0,1}$ is
the first zero of $J_{0}(z)$ and the higher eigenvalues are
similarly expressed through other Bessel functions zeros. The
integral density of states corresponding to the spectrum of $t$ is
given by
\begin{equation}\label{IntDensStates}
  {\cal N}(\varepsilon)= \left[2\sqrt{\pi\,\Gamma(3/2)} \right]^{-1}
  \sum_{j=1}^{\infty}\theta(\varepsilon-E_{j})\sqrt{\varepsilon-E_{j}}\,,
\end{equation}
where $\theta$ is Heaviside function. This gives for the PBG in
the infinite tube the grand canonical total particle density
\begin{equation}\label{dens-1}
\rho(\beta ,\mu )=\int_{0}^{\infty} {\cal
N}(d\varepsilon)\frac{1}{ e^{\beta (\varepsilon-\mu )}-1}
\end{equation}
for temperature $\beta^{-1}\ge0$ and chemical potential $\mu
<E_1$. Since the critical value $\rho _{c}(\beta ):=\lim_{\mu
\nearrow E_1} \rho (\beta,\mu)$ of the density is infinite, there
is no BEC of the PBG in this quasi-one-dimensional system.

Several possibilities are known to make the critical density
finite, provoking thus a BEC, by changing the one-particle
spectrum. One is to replace the Dirichlet boundary condition by a
"sticky" one, i.e. by a mixed condition with a positive outside
gradient of the one particle wave-function on the boundary
\cite{VVZ}. One can also switch in an external local attractive
potential producing bound state(s) below the $\inf\sigma(t) =
E_1$, the continuum spectrum \cite{LVZ}. A less obvious way is a
suppression of the density states at the bottom of the spectrum,
leading to convergence of the integral (\ref{dens-1}) for
$\mu=E_1$, by embedding the PBG into a random external potential
\cite{LPZ}.

In the present note we are going to show that a
\textit{geometrical deformation} of the straight tube such as a
simple local bending, even a gentle one, may produce the BEC of
the PBG with the condensate localized in the vicinity of such a
bend. We will also discuss other types of (local) geometrical
deformations creating the BEC in these quasi-one dimensional
systems as well the conditions under which the effect could be
experimentally attainable.

To this aim we have to recall the known results about one-particle
spectra in deformed tubes. Consider first a local bending of the
infinite cylinder ${\cal C}(S_r)$, so that the tube axis is a
smooth curve which is straight outside a compact region; we also
suppose that the bent tube ${\cal C}^*(S_r)$ does not intersect
itself. If the deformation is nontrivial, ${\cal C}^*(S_r) \ne
{\cal C}(S_r)$, it generates one or more eigenvalues
\cite{GJ92,DE95} below the continuum threshold $E_1$.

Their number is finite and the lowest one of them is simple; the
form of this discrete spectrum is determined the geometry of the
deformed tube. Properties of such bound states in bent tubes
are well understood: \\
(a1) If the tube is only slightly bent there is only bound state
with energy $\epsilon^*_1$ and the gap $E_1-\epsilon^*_1$ is
proportional to $\varphi^4$, where $\varphi$ is the bending angle
of ${\cal C}^*\,$, see \cite{DE95}. \\
(a2) The tube need not be straight outside a bounded region; it is
sufficient that it is asymptotically straight in the sense that
the tube axis curvature decays faster than $|s|^{-1}$ where $s$ is
the arc length of the tube axis. \\
(a3) The cross section need not be circular. In such a case,
however, the shape is restricted by the so-called \emph{Tang
condition} imposed on the torsion \cite{DE95}. It is satisfied, in
particular, if ${\cal C}^*= {\cal S}^*\times [0,d]$, where
${\cal S}^*$ is a bent planar strip \cite{ES}. \\
(a4) The effect is robust. It also does not require the tube
${\cal C}^*$ to be bent smoothly; the geometrically induced bound
states exist also in sharply bent tubes \cite{sharp}.

Furthermore, bending is not the only way how to produce
geometrically a nonempty discrete spectrum: \\
(b1) Another mechanism is a local change of the cross section.
Protrusions and other deformations which do not reduce the volume
also give rise to an effective attractive interaction \cite{BGRS97}. \\
(b2) Similar effect comes from a tube branching. A cross-form tube
is known to support a single bound state \cite{SRW89}, numerous
isolated eigenvalues arise in a skewed scissor-shaped cross with
small enough angle \cite{BEPS02}. \\
(b3) One more example are two parallel tubes with a window in the
common boundary, where the number of bound states is given by the
window length \cite{ESTV96}.

The geometrically induced discrete spectrum is finite in
asymptotically straight ducts. In the bent-tube case, for
instance, it follows from the fact that the effective attractive
potential falls off faster that $|s|^{-2}$, see \cite{DE95}. On
the other hand, the actual form of the spectrum depends on tube
geometry. In particular, if we deal with two or more well
localized, identical, and mutually distant perturbations the
eigenvalues cluster around those of a single deformation, with a
split exponentially small w.r.t. the distance between the
perturbations. This is proved mathematically for a tube with a
pair of windows \cite{BE04}, however, various examples worked out
numerically \cite{LCM} suggest that such a behavior is generic.

Consider thus such a geometrically induced discrete spectrum below
the continuum threshold \cite{threshold} in a deformed cylinder
which consists consists of one or several bound states with
energies $\left\{\epsilon_{s}^* < E_{1}:\: s=1,2,\dots \,\right\}$
and eigenfunctions localized in the vicinity of the
deformation(s). If one orders the bound-state energies naturally,
$\epsilon_{1}^* < \epsilon_{2}^* \leq \ldots < E_{1}$, then the
domain of the allowed chemical potentials is given by the
inequality $\mu\leq\epsilon_{1}^*$. This implies that the critical
particle density is bounded,
\begin{equation}\label{crit-inf-2}
\rho^*_{c}(\beta ):=\lim_{\mu \nearrow
\varepsilon_{1}^*}\int_{E_{1}}^{\infty }{\cal
N}^{*}(d\varepsilon)\frac{1}{ e^{\beta (\varepsilon-\mu )}-1}<
\infty\,;
\end{equation}
here ${\cal N}^{*}(\varepsilon)$ is the integrated density of
states for the deformed cylinder. This opens a way to create the
BEC.

To make this effect transparent consider first a finite segment of
the deformed cylinder ${\cal C}^{*}_{L}$ of length $L$. Then the
Dirichlet Laplacian which plays here the role of the one-particle
Hamiltonian,
\begin{equation}\label{Lap-L}
  t_{*}(L) := -\Delta_D^{{\cal C}^{*}_L}\,,\quad
  {\rm Dom}(t_{*}(L)) = W_{0}^{2,2}({\cal C}^{*}_L)\,,
\end{equation}
has a purely discrete spectrum $\sigma(t_{*}(L))$ accumulating at
infinity. Denote by $P_{I}(t_{*}(L))$ the spectral projection of
$t_{*}(L)$ for a Borel set $I\subset \mathbf{R}$. It allows us to
write the finite-volume integrated density of states corresponding
to the operator (\ref{Lap-L}) as
\begin{equation}\label{L-IntDensStat}
{\cal N}^{*}_{L}(\epsilon):= \frac{1}{\left|{\cal C}^{*}_L\right|}
\,\mbox{Tr}\, \left\{P_{(-\infty,\epsilon)}(t_{*}(L))\right\},
\end{equation}
where $\left|{\cal C}^{*}_L\right|$ is the volume of the segment
${\cal C}^{*}_L$. By (\ref{L-IntDensStat}) the total particle
density $\rho^*_{L}(\beta ,\mu )$ in ${\cal C}^{*}_L$ acquires the
form
\begin{eqnarray}\label{L-PartDens}
\lefteqn{\int_{-\infty}^{\infty }{\cal
N}^{*}_{L}(d\varepsilon)\frac{1}{ e^{\beta (\varepsilon-\mu )}-1}
 =\frac{1}{\left|{\cal
C}^{*}_L\right|}\sum_{\left\{\epsilon^{*}_{s}(L)\right\}}\frac{1}{
e^{\beta (\epsilon^{*}_{s}(L)-\mu )}-1}} \nonumber \\ &&
\phantom{AA} +\frac{1}{\left|{\cal
C}^{*}_L\right|}\sum_{\left\{\varepsilon^{*}_{j}(L)\right\}}\frac{1}{
e^{\beta (\varepsilon^{*}_{j}(L)-\mu )}-1}\,. \phantom{AAAAAAAAA}
\end{eqnarray}
Here $\{\epsilon^{*}_{s}(L)\}_{s\geq1}$ and
$\{\varepsilon^{*}_{j}(L)\}_{j\geq1}$ are eigenvalues of the
operator $t_{*}(L)$ divided into two groups with
$\epsilon^{*}_{s}< \varepsilon^{*}_{j}$ for any $s,j$. The first
one consists of those which converge to the eigenvalues of the
infinite-tube operator $t_{*}$ as $L\to\infty$; recall that all of
them are monotonously decreasing with respect to $L$. The limit is
naturally taken in such a way that the distance of the cut-offs
from the deformed part(s) tend to infinity. On the other hand,
$\{\varepsilon^{*}_{j}(L)\}_{j\geq1}$ are those eigenvalues which
give in this limit the continuous spectrum of $t_{*}$. Combining
this with the above stated properties of the one-particle spectrum
we see that the first group $\{\epsilon^{*}_{s}(L)\}_{s\geq1}$ is
finite. Since $\left|{\cal C}^{*}_L\right| \to \infty$ as
$L\to\infty$, one gets then from (\ref{L-PartDens}) for the
limiting density of states $\rho^*(\beta,\mu ):=
\lim_{L\to\infty}\rho_{L}^*(\beta,\mu)$ the relation
\begin{eqnarray}\label{Lim-mu-neg}
\rho^*(\beta ,\mu ) &=&
\lim_{L\rightarrow\infty}\frac{1}{\left|{\cal C}^{*}_L\right|}
\sum_{\left\{\varepsilon^{*}_{j}(L)\right\}}\frac{1}{ e^{\beta
(\varepsilon^{*}_{j}(L)-\mu )}-1} \nonumber \\
&=&\int_{E_{1}}^{\infty }{\cal N}^{*}(d\varepsilon)\frac{1}{
e^{\beta (\varepsilon-\mu )}-1}
\end{eqnarray}
provided $\mu \leq \lim_{L\rightarrow\infty}\epsilon^{*}_{1}(L)=
\epsilon^{*}_{1} < E_{1}$, uniformly in $\mu$, where ${\cal
N}^{*}(d\varepsilon)$ is the weak limit of the measure family
$\left\{{\cal N}^{*}_{L}(d\varepsilon)\right\}_L$. In particular,
the limit (\ref{Lim-mu-neg}) implies
\begin{equation}\label{ro-c-*}
\rho^*(\beta ,\mu)< \rho_{c}^{*}(\beta): = \rho^*(\beta ,
\epsilon^{*}_{1})
\end{equation}
for $\mu < \epsilon^{*}_{1}$ in view of (\ref{crit-inf-2}).

Now let the total particle density be $\rho >
\rho^{*}_{c}(\beta)$. To show that this implies the BEC, let us
analyze solutions $\left\{\mu_{L}(\beta,\rho)\right\}_{L}$ of the
equation following from (\ref{L-PartDens}),
\begin{equation}\label{mu-equa}
    \rho = \rho_{L}^*(\beta ,\mu )\,;
\end{equation}
notice that they always exist and are bounded by
$\epsilon^{*}_{1}(L)$ from above. Using (\ref{L-PartDens}) and
(\ref{Lim-mu-neg}) one gets for $\rho \leq \rho_{c}^*(\beta)$
\begin{equation}\label{ro<ro-c}
\lim_{L\rightarrow\infty}\mu_{L}(\beta,\rho)= \mu(\beta,\rho) \leq
\epsilon^{*}_{1}\,.
\end{equation}
On the other hand, for $\rho > \rho^{*}_{c}(\beta)$ we rewrite
eq.~(\ref{mu-equa}) as
\begin{eqnarray}\label{eq-ro>ro-c}
\lefteqn{ \rho - \frac{1}{\left|{\cal
C}^{*}_L\right|}\sum_{\left\{\varepsilon^{*}_{j}(L)\right\}}\frac{1}{
e^{\beta (\varepsilon^{*}_{j}(L)- \mu_{L}(\beta,\rho))}-1} }
\nonumber \\ && \phantom{A} = \frac{1}{\left|{\cal
C}^{*}_L\right|}\sum_{\left\{\epsilon^{*}_{s}(L)\right\}}\frac{1}{
e^{\beta (\epsilon^{*}_{s}(L)-\mu_{L}(\beta,\rho))}-1}\,.
\end{eqnarray}
(i) Suppose that the bound-state energies
$\left\{\epsilon^{*}_{s}(L)\right\}_{s\geq1}$ verify the following
conditions as $L\to\infty$,
\[E_1 - \delta > \epsilon^{*}_{1}(L)\quad\mbox{for some}\; \delta > 0\]
and
\[\epsilon^{*}_{s}(L)-
\epsilon^{*}_{1}(L)\geq a\left|{\cal C}^{*}_L\right|^{\alpha-1}
,\quad a>0\,,\:\alpha>0 \,.\]
This is true, in particular, for a tube with a single bend or
protrusion and two cut-offs moving away of it, where in view of
the norm-resolvent convergence of $-\Delta_D^{{\cal C}^{*}_L}$ to
$-\Delta_D^{{\cal C}^{*}}$ the eigenvalue $\epsilon^{*}_{1}(L)$
tends to $\epsilon^{*}_{1}<E_1$ and the eigenvalue difference to a
nonzero limit.

Since $\mu_{L}(\beta,\rho)< \epsilon^{*}_{1}(L)$, we can use the
uniform convergence (\ref{Lim-mu-neg}) to obtain the asymptotics
of the solution of eq. (\ref{eq-ro>ro-c}) as $L\to\infty$, namely
\begin{equation}\label{mu-solut}
\mu_{L}(\beta,\rho)= \epsilon^{*}_{1}(L) - \frac{1}{\beta(\rho
\!-\! \rho_{c}^{*}(\beta))\left|{\cal C}^{*}_L\right|} +
o\left(\left|{\cal C}^{*}_L\right|^{-1}\right).
\end{equation}
This means that $\lim_{L\rightarrow\infty}\mu_{L}(\beta,\rho)=
\epsilon^{*}_{1}$, and since $\left|{\cal C}^{*}_L\right|= {\cal
O}(L)$ as $L\to\infty$, the thermodynamic limit in
(\ref{eq-ro>ro-c}) gives rise to BEC at the lowest level
$\epsilon^{*}_{1}$ only,
\begin{equation}\label{1-levelBEC}
\rho - \rho_{c}^{*}(\beta)=
\lim_{L\rightarrow\infty}\frac{1}{\left|{\cal
C}^{*}_L\right|}\frac{1}{ e^{\beta
(\epsilon^{*}_{1}(L)-\mu_{L}(\beta,\rho))}-1}\
\end{equation}
The same is true for other tube geometries, for instance, for a
tube with several well distinguished bends or ``bubbles'', as long
as the limit $L\to\infty$ means that cut-offs move away in the
asymptotically straight parts.

(ii) The situation may change when the thermodynamic limit
$L\rightarrow\infty$ is more complicated and involves a local
change of the geometry as well. As a model example, consider again
a tube with a finite number $n>1$ of well distinguished
\emph{identical} bends, but suppose now that the distances between
them increase also with increasing $L$. As we have recalled above,
the first $n$ eigenvalues $\left\{\epsilon^{*}_{s}
(L)\right\}_{s=1}^{n}$ then cluster being exponentially close to
each other with respect to the separation parameter $L$,
\begin{equation}\label{exp-deg}
\epsilon^{*}_{s}(L)- \epsilon^{*}_{1}(L)\leq C\, e^{-a L}\,,\quad
1\leq s \leq n\,,
\end{equation}
for some positive $C,a$. By (\ref{exp-deg}) the asymptotics of the
solution to eq.~(\ref{eq-ro>ro-c}) is again of the form
(\ref{mu-solut}). This means that the limit in (\ref{eq-ro>ro-c})
gives rise to BEC equally fragmented into the group of $n$ levels
$\left\{\epsilon^{*}_{s}(L)\right\}_{s=1}^{n}$ which are almost
degenerate being exponentially separated,
\begin{equation}\label{n-levelBEC}
\lim_{L\rightarrow\infty}\frac{1}{\left|{\cal
C}^{*}_L\right|}\frac{1}{ e^{\beta
(\epsilon^{*}_{s}(L)-\mu_{L}(\beta,\rho))}-1}=\frac{\rho -
\rho_{c}^{*}(\beta)}{n}
\end{equation}
for $1\leq s \leq n$. This fragmentation is called a
\textit{type-I generalized} BEC , contrasting with the case of the
\textit{infinite} fragmentation known as the \textit{type-II
generalized} BEC, see \cite{ZB}, \cite{BL} and also \cite{type3}.
What is important is that this condensate is separated from the
continuum spectrum by a finite energy gap which makes it more
stable than the conventional BEC, see the discussion in
\cite{LVZ}.

To analyze localization properties of the geometrically induced
BEC we employ the PBG \textit{one-body reduced density matrix}
with the kernel $\rho _{L}(\beta ,\mu ;x,y)$ given by
\begin{equation}\label{L-RDM}
\left|{\cal C}^{*}_L\right|\int_{-\infty}^{\infty }{\cal
N}^{*}_{L}(d\varepsilon)\frac{1}{e^{\beta (\varepsilon-\mu )}-1
}\: \overline{\psi^{*}_{\varepsilon,L}(x)}
\psi^{*}_{\varepsilon,L}(y)\,,
\end{equation}
where $\left\{\psi^{*}_{\varepsilon,L}\right\}$ are the
\textit{normalized eigenfunctions} of the operator (\ref{Lap-L}).
The diagonal part of the matrix (\ref{L-RDM}) is the
\textit{local} particle density, since by (\ref{L-PartDens})
\begin{equation}\label{tot-numb}
\int_{{\cal C}^{*}_L} dx \rho _{L}(\beta ,\mu ;x,x)= \left|{\cal
C}^{*}_L\right|\rho_{L}(\beta ,\mu )
\end{equation}
is the total number of particles in ${\cal C}^{*}_L$. In fact a
relevant quantity for the BEC \textit{space localization} is the
\textit{local} particle density \emph{per unit volume}
$\tilde{\rho}_{L}(\beta ,\mu ;x)$ given by
\begin{equation}\label{LPDpUV}
\int_{-\infty}^{\infty }{\cal
N}^{*}_{L}(d\varepsilon)\frac{1}{e^{\beta (\varepsilon-\mu )}-1
}\:\overline{\psi^{*}_{\varepsilon,L}(x)}
\psi^{*}_{\varepsilon,L}(x)\,.
\end{equation}
Indeed, since in the limit only the eigenfunctions corresponding
the eigenfunction family $\left\{\psi^{*}_{s}\right\}_{s\geq1}
\subset L^{2}({\cal C}^{*})$ related to the infinite-tube bound
states is preserved, in contrast to all the others which are
extended, we get
\begin{eqnarray}\label{BEC-loc-1}
\tilde{\rho}(\beta ,\mu ;x) &\!:=\!&
\lim_{L\rightarrow\infty}\int_{-\infty}^{\infty }{\cal
N}^{*}_{L}(d\varepsilon)\frac{1}{e^{\beta (\varepsilon-\mu )}-1
}\: \overline{\psi^{*}_{\varepsilon,L}(x)}
\psi^{*}_{\varepsilon,L}(x) \nonumber \\ &\!=\!& (\rho -
\rho_{c}^{*}(\beta))\:\overline{\psi^{*}_{\epsilon_1}(x)}
\psi^{*}_{\epsilon_1}(x)
\end{eqnarray}
for $\rho > \rho_{c}^{*}(\beta)$ in the case (\ref{1-levelBEC}).
For the fragmented BEC (\ref{n-levelBEC}) one gets for $\rho >
\rho_{c}^{*}(\beta)$ similarly
\begin{equation}\label{BEC-loc-few}
\tilde{\rho}(\beta ,\mu ;x)=\frac{\rho -
\rho_{c}^{*}(\beta)}{n}\sum_{s=1}^{n}\left|
\psi^{*}_{\epsilon_s}(x)\right|^2;
\end{equation}
it is obvious that $\tilde{\rho}(\beta ,\mu ;x) =0$ holds when
$\rho \leq \rho_{c}^{*}(\beta)$.

Thus in contrast to the BEC in the translation-invariant case, the
$L^2$-localized condensation in the bounded states correspond to
an \emph{infinite} accumulation of the local particle density
defined by (\ref{L-RDM}). On the other hand, by (\ref{1-levelBEC})
and (\ref{n-levelBEC}) in combination with (\ref{BEC-loc-1}),
(\ref{BEC-loc-few}), the \textit{total} density of particles
condensed in the bounded states,
\begin{equation}\label{BEC-dens}
\rho - \rho_{c}^{*}(\beta)= \int_{{\cal C}^{*}} dx
\tilde{\rho}(\beta ,\mu ;x)\,,
\end{equation}
is \emph{finite}.

After this theoretical analysis let us ask about chances to
observe the described type of the BE condensation in an
experiment. We are not going to discuss technically the ways in
which the Bose gas can be confined in a geometrically deformed
tube, generally we have in mind either a modification of the
existing elongated traps \cite{cigar} or using hollow optical
fibers \cite{fiber} as suggested in \cite{EV99}.

The important parameter is the tube radius which for both the
elongated traps and hollow fibers can be made as small as
$r\approx 5\,\mu m$ which determine the threshold energy by the
mentioned expression $E_1= \frac{\hbar^2}{2M}\, j_{0,1}^2 r^{-2}$,
where $M$ is atom of the mass in question. Let us further
introduce the \emph{relative gap} size by $\gamma_{\rm rel}:=
(E_1-\epsilon_1^*)/E_1$. This quantity can theoretically reach the
value about $0.39\,$ in a smoothly bent tube \cite{EFK04}, but a
typical value for a bending angle of $90^{\rm o}$ and more is
around $\gamma_{\rm rel}\approx 10^{-1}$, see \cite{LCM}.

Two conditions must be satisfied. First of all, the
bending-induced gap $\gamma_{\rm rel}E_1$ must be much larger than
the effect of the finite length $L$ of the recipient. The latter
is characterized by the first longitudinal eigenvalue
$\frac{\hbar^2} {2M}\, (\pi/L)^2$. It is clear that the trap must
sufficiently elongated to fulfill the condition, roughly speaking
$L/r \gtrsim 20$ is sufficient. This is true in both situations
mentioned above, for hollow fibers it can be even much better.

More complicated is the question of thermal stability; the said
gap must be larger than the energy $k_{\rm B}T$. This determines a
critical temperature above which the bending-induced BEC is likely
to be destroyed by thermal fluctuations. Using the value $r\approx
5\,\mu m$ we find that
\begin{equation}\label{Tcrit}
T_{\rm crit} \approx 5.4\,\times\,10^{-8} \:\frac{\gamma_{\rm
rel}}{Z}\,,
\end{equation}
where $Z$ is the atomic number in question. Hence lighter nuclei
are preferable; there is one order of magnitude difference between
${\rm Li^7}$ and ${\rm Ru^{87}}$. With above estimate in mind,
however, even for the light ones it is difficult to perform the
measurement in the available nanokelvin conditions. On the other
hand, the effect is not too far from the experimental reach;
recall for instance that squeezing the transverse size to a one
micron radius would enhance the critical temperature (\ref{Tcrit})
by the factor of 25.

Let us finally mention one more feature of such a geometrically
induced BE condensate. The ground-state wave function into which
will the atoms of the Bose gas condensate -- cf. (\ref{BEC-loc-1})
-- is exponentially localized away of the bend; for examples of
such wave functions see \cite{LCM} and the literature mentioned
there. This means that, e.g., bending a tube would mean not only
the condensation but also that the condensate will squeezed in the
bend the more the larger the bending angle is.

In conclusion, we have demonstrated here a purely geometric way to
achieve a stable BEC based on local deformations of a tube-shaped
recipient and we discussed ways in which the effect could be
experimentally observed.


\acknowledgements This work was supported by ASCR within the
project K1010104 and the ASCR-CNRS exchange program.


\begin{thebibliography}{99}
\bibitem{threshold}
The threshold $\inf \sigma(t_*) = E_1$ is preserved as long as the
tube remains to be asymptotically straight
\end{thebibliography}
\end{document}